\documentclass[aip, jcp, citeautoscript, amsmath, amssymb, reprint,
]{revtex4-2}
\usepackage[title]{appendix}
\usepackage{graphicx, tikz, orcidlink}
\usepackage{epstopdf}
\usepackage{ascmac}
\usepackage{ulem}
\usepackage{cancel}
\usepackage{here}
\usepackage{physics}
\usepackage{bm}
\usepackage{comment}
\usepackage{booktabs}
\usepackage[inline]{enumitem}
\usepackage{listings}
\usepackage{xcolor}
\usepackage{multirow}
\usepackage{amsthm,color}
\newtheoremstyle{query}%
{}{}%space above/below
{\color{red}}%body style
{}%heading indent
{\sffamily\bfseries}{:}{12pt}%heading style/punctuation/space after
{}% head spec
\theoremstyle{query}
\newtheorem{aq}{Author Query/Comment}

\newcommand{\baq}{\begin{aq}}%This just makes things easier
\newcommand{\eaq}{\end{aq}}

\begin{document}
\title{sbml4md: A computational platform for 
System-Bath Modeling via Molecular Dynamics powered by Machine Learning}
\date{Last updated: \today}

\author{Kwanghee Park\orcidlink{0009-0005-5223-8234}}
\email[Author to whom correspondence should be addressed: ]{park.kwanghee.v48@kyoto-u.jp}
\affiliation{Department of Chemistry, Graduate School of Science,
Kyoto University, Kyoto 606-8502, Japan}\author{Seiji Ueno\orcidlink{0000-0002-7050-1084}}
\affiliation{HPC Systems Inc., Japan}
\altaffiliation{Department of Chemistry, Kyoto University, Kyoto 606-8502, Japan}
\author{Yoshitaka Tanimura\orcidlink{0000-0002-7913-054X}}
\email[Author to whom correspondence should be addressed: ]{tanimura.yoshitaka.5w@kyoto-u.jp}
\affiliation{Department of Chemistry, Graduate School of Science,
Kyoto University, Kyoto 606-8502, Japan}

\begin{abstract}
We introduce sbml4md, a newly developed algorithm implemented as a software package to extract parameters of multimode anharmonic Brownian (MAB) models from molecular dynamics (MD) trajectories for simulating nonlinear vibrational spectra of intramolecular modes of molecular liquids. By leveraging machine learning (ML) techniques to capture vibrational anharmonicity, intermolecular couplings, and bath correlation functions for each mode, sbml4md obviates empirical fitting and enables the modeling of environments with spatial and temporal heterogeneity. This work provides a set of parameters specifically tailored for the Hierarchical Equations of Motion (HEOM) framework, enabling numerically “exact” simulations of nonlinear vibrational spectra. Building upon our previous implementation for intramolecular vibrational modes [Park, Jo, and Tanimura, J. Chem. Phys. 163, 214104 (2025)], the present code enhances optimization efficiency by explicitly accounting for intermolecular vibrational contributions. This extension enables sbml4md to broaden the applicability of HEOM-based dynamical modeling by seamlessly integrating classical MD approaches, thereby providing a flexible and scalable framework for simulating both linear and nonlinear spectra under realistic conditions with minimal empirical input. The accompanying ML code, written in Python, is provided as supporting material.
\end{abstract}
\maketitle

\section{Introduction}
\label{sec:intro}

Modern laser experiments, particularly ultrafast nonlinear spectroscopies, exhibit pronounced sensitivity to quantum coherence in complex molecular systems excited by femtosecond pulses.\cite{Cho2009,Hamm2011ConceptsAM,HammPerspH2O2017} These experimental advances necessitate theoretical simulations and analyses to interpret the observed dynamics and spectral features.\cite{mukamel1999principles,TI09ACR,JansenChoShinji2DVPerspe2019}

Recent progress in molecular dynamics (MD) simulation has enabled the exploration of a wide range of chemical and biological phenomena, including systems comprising millions to hundreds of millions of molecules with remarkable accuracy.\cite{Shinji2DRaman2006,HT06JCP,LHDHT08JCP,HT08JCP,Wei2015Nagata2DRamanTHz,IHT14JCP,IHT16JPCL,JianLiu2018H2OMP,Imoto_JCP135,Paesan2018H2OCMD}
Nevertheless, theoretical modeling remains indispensable for analyzing nonlinear spectroscopic signals.\cite{ChoOhmine1994,mukamel1999principles,TI09ACR,JansenChoShinji2DVPerspe2019,Mukamel2009,SkinnerStochs2003,Skinner2005,JansenSkinnerJCP2010} This requirement stems from the inherently quantum nature of spectroscopic processes, which elude adequate description by classical dynamics and thus demand rigorous quantum mechanical treatment.\cite{SkinnerCPL2004,IT08CP,TI09ACR} The challenge is further compounded by the complexity of molecular behavior, such as anharmonic mode coupling and thermally driven excitation and relaxation. 
A model that accurately reproduces both the peak positions and spectral profiles of nonlinear response functions provides compelling evidence that it captures the essential quantum dynamical characteristics of the system, thereby enabling meaningful interpretation.

To date, the construction of such models has relied heavily on physical intuition, with model parameters fitted to experimental and simulation data. While this approach can capture salient features of complex systems, its reliability is often insufficiently tested. For instance, linear absorption spectra---governed by linear response theory---are relatively insensitive to anharmonicity and quantum coherence. To evaluate models more rigorously, nonlinear response functions serve as sensitive and informative benchmarks.\cite{mukamel1999principles,TI09ACR}

Our group has conducted multidimensional spectral analyses based on the multimode anharmonic Brownian (MAB) model.\cite{IIT15JCP,IT16JCP,TT23JCP1,TT23JCP2,ST11JPCA,HT25JCP1,HT25JCP2} Initially, model parameters were tuned to reproduce two-dimensional vibrational spectra (2DVS) derived from molecular dynamics simulations and experiments. The MAB model has proven highly effective in describing the features of 2DVS.\cite{IIT15JCP,IT16JCP} With the emergence of machine learning (ML), it has become possible to construct models directly from large-scale datasets without relying on empirical fitting. We employed ML techniques to extract MAB model parameters from the MD trajectories, including the anharmonicity of each intramolecular mode, the strength of anharmonic mode–mode couplings, the linear and nonlinear system–bath coupling strengths, and the spectral distribution function (SDF) of the bath.\cite{UT20JCTC,PJT25JCP1} While physical intuition still guides many modeling efforts, beginning with a microscopic foundation and refining it through ML enables systematic validation of model assumptions. In this work, we enhance our previous algorithm,\cite{UT20JCTC} implement it as open-source software, and provide documentation and illustrative examples.

This paper is organized as follows. Section~\ref{sec:theory} introduces the MAB model for intramolecular vibrational modes and presents its extension to account for the effects of intermolecular vibrational modes. The classical hierarchical Fokker–Planck equations for a three-mode MAB model in 2DVS (hereafter referred to as QHFPE-2DVS) are also presented in Appendix \ref{HEOM_Drude}. Section~\ref{sec:software-architecture} describes the machine learning algorithm implemented in \texttt{sbml4md} to extract model parameters from molecular dynamics simulations. Section~\ref{sec:IOSpecification} outlines the input/output specifications. 
Section~\ref{sec:Demo} presents the results and analysis of model parameters associated with three intramolecular vibrational modes of liquid water, obtained using two different MD force fields.  Section~\ref{sec:conclusion} concludes this study.

\section{MAB Model}
\label{sec:theory}

Consider a situation in which a target molecule---whether a solute in a liquid solvent or a reactive species embedded in a biological or molecular solid matrix---interacts with its surrounding environment.
To simulate both linear and nonlinear vibrational spectra of the molecule in such a situation, we employ the MAB model expressed as\cite{IIT15JCP,IT16JCP,TT23JCP1,TT23JCP2,ST11JPCA,HT25JCP1,HT25JCP2} 
\begin{eqnarray}
  \hat{H}_\mathrm{tot}&&= \sum_{s} \left( \hat{H}_{A}^{(s)} + \sum_{s>s'} \hat{U}_{ss'}(\hat{q}_s, \hat{q}_{s'}) \right)  \nonumber  \\
    &&+ \sum_{s} \sum_{j_s} \left[ \frac{\hat{p}_{j_s}^{2}}{2m_{j_s}} + \frac{m_{j_s}\omega_{j_s}^{2}}{2} \left( \hat{x}_{j_s} - \frac{\alpha_{j_s} \hat{V}_s(\hat{q}_s)}{m_{j_s}\omega_{j_s}^2} \right)^2 \right] \nonumber  \\
    &&+ \hat{H}_\mathrm{inter}^{\rm PBM}.
  \label{eqn:H_total}
\end{eqnarray}
Here, the Hamiltonian for the $s$th mode of the targeting molecule
is defined as
\begin{eqnarray}
\hat{H}_{A}^{(s)} = \frac{\hat{p}_s^{2}}{2m_s} + \hat{U}_s(\hat{q}_s),
\label{eqn:SH}
\end{eqnarray}
where $m_s$ is the mass, $\hat{q}_s$ the coordinate, and $\hat{p}_s$ the momentum of the $s$th mode. The anharmonic potential for the $s$th mode is given by
\begin{eqnarray}
\hat{U}_s(\hat{q}_s) = \frac{1}{2} m_s \omega_s^2 \hat{q}_s^2 + \frac{1}{3!} g_{s^3} \hat{q}_s^3,
\label{eqn:Potential_s}
\end{eqnarray}
where $\omega_s$ denotes the vibrational frequency and $g_{s^3}$ represents the cubic anharmonicity. 
The interaction potential between the $s$th and $s'$th vibrational modes is formulated as
\begin{eqnarray}
\hat{U}_{ss'}(\hat{q}_s, \hat{q}_{s'}) &&= g_{s{s'}}\hat{q}_s\hat{q}_{s'} \nonumber \\
&&+ \frac{1}{6}  \qty(g_{s^2s'}\hat{q}_s^2 \hat{q}_{s'} + g_{s{s'}^2} \hat{q}_s \hat{q}_{s'}^2 ),
\label{eqn: Potential ss'}
\end{eqnarray}
where the set $\{g_{ss'},\, g_{s^{2}s'},\, g_{s{s'}^{2}}\}$ represents the mode-mode coupling parameters. The coefficient $g_{ss'}$ denotes the second-order anharmonic coupling, whereas $g_{s^{2}s'}$ and $g_{s{s'}^{2}}$ characterize the third-order contributions. The cubic cross‑coupling terms induce the Fermi resonance, thereby affecting frequency shifts and energy redistribution on the time scales relevant to nonlinear response functions. The environmental influence on the $s$-mode is modeled by a bath composed of harmonic oscillators indexed by $j_{s}$. Each oscillator is characterized by its momentum $p_{j_{s}}$, coordinate $x_{j_{s}}$, mass $m_{j_{s}}$, frequency $\omega_{j_{s}}$, and coupling strength $\alpha_{j_{s}}$, which specifies the interaction between the bath oscillator and the molecular vibration of the $s$-mode.
To preserve the system's translational invariance, a counter term is incorporated into each bath, as illustrated in Refs.~\onlinecite{TW91PRA,OT97PRE}.

In this study, we examine how the intramolecular $s$ mode of a single molecule interacts with surrounding molecules---such as a water molecule in liquid water---through intermolecular interactions. To describe this process, we employ a model in which the anharmonic oscillator associated with the $s$ mode interacts nonlinearly with a bath.
Although 2DVS has clarified the roles of vibrational relaxation and dephasing as central mechanisms governing molecular motion,\cite{Skinner2002HOD1,Skinner2002HOD2,Skinner2003HOD3,Skinner2003HOD4,SkinnerStochs2003,SkinnerVI,SkinnerVII,YagasakiSaitoJCP20082DIR,YagasakiSaitoJCP2011Relax,Yagasaki_ARPC64,ImotXanteasSaitoJCP2013H2O,Imotobend-lib2015} it remains crucial to incorporate non-Markovian S–B interactions---particularly those of the linear–linear (LL) and square–linear (SL) types---alongside anharmonic mode–mode couplings.\cite{KT02JCP1,KT04JCP,IT06JCP,OT97PRE} Accordingly, we express the system component of the system-bath (S–B) interaction, $\hat{V}_{s}({\hat{q}_s})$,
as a combination of LL and SL contributions:
\begin{eqnarray}
  \hat{V}_{s}(\hat{q}_s)\equiv \hat{V}^{(s)}_{\mathrm{LL}}\hat{q}_s+ \frac{1}{2} \hat{V}^{(s)}_{\mathrm{SL}}\hat{q}_s^{2},
  \label{eqn: LLSL}
\end{eqnarray}
where $V^{(s)}_{\mathrm{LL}}$ and $V^{(s)}_{\mathrm{SL}}$ denote the respective coupling strengths.\cite{T06JPSJ,TI09ACR}

To characterize the properties of environment, we employ the SDF defined as
\begin{eqnarray}
J_{s}(\omega)= \sum_{j_n} \frac{\alpha_{j_s}^2}{2m_{j_s} \omega_{j_s}} \delta(\omega - \omega_{j_s})
\end{eqnarray}
together with the inverse temperature $\beta=1/k_{\mathrm{B}}T$, where $k_{\mathrm{B}}$ is Boltzmann's constant and $T$ is the thermodynamic temperature. 
To apply the HEOM formalism, we adopt the Drude SDF,
  \begin{eqnarray}
  J_s(\omega) = \frac{m_s \zeta_s}{2\pi} \frac{\gamma_s^2 \omega}{\omega^2 + \gamma_s^2},
  \label{eq:drude}
  \end{eqnarray}
where $\zeta_{s}$ and $\gamma_{s}$ represent the S-B coupling strength and the inverse correlation time of the bath fluctuations, respectively.

The bath introduced in Eq.~\eqref{eqn:H_total} is, in principle, required to account for all thermal effects, including those arising from intermolecular modes.
However, during the recent ML optimization of the MAB model to determine the SDF in Eq.~\eqref{eq:drude} from MD trajectories,\cite{PJT25JCP1} 
we found that the optimization efficiency is markedly enhanced by introducing an additional Brownian oscillator (BO) bath mode on top of the baths described above. 
Importantly, the optimized parameters of the SDF in Eq.~\eqref{eq:drude} remain unchanged irrespective of whether this BO bath mode is included, 
even though its presence produces unphysical artifacts in the linear spectrum. 
This discrepancy arises because trajectory optimization should specifically treat intermolecular modes independently, whereas the Drude SDF-based scheme inevitably channels energy exchange through intramolecular modes.

For this reason, we employ the BO mode solely as an independent dummy bath to facilitate trajectory optimization, 
while omitting it from intramolecular spectral calculations. 
Despite its auxiliary role, the BO mode effectively captures essential intermolecular characteristics and thus serves as a pseudo Brownian mode (PBM).

The effective Hamiltonian representing the intermolecular PBMs is constructed using the functions introduced above, as follows:
\begin{eqnarray}
&& \hat{H}_\mathrm{inter}^{\rm PBM}= \sum_{\alpha} \qty( \hat{H}_{A}^{(\alpha)} +  \sum_{s}  \hat{U}_{s\alpha}(\hat{q}_s, \hat{q}_{\alpha}))   \nonumber  \\
   &&~~~~+ \sum_{\alpha} \sum_{j_{\alpha}}\qty[\frac{\hat{p}_{j_{\alpha}}^{2}}{2m_{j_{\alpha}}}+\frac{m_{j_{\alpha}}\omega_{j_{\alpha}}^{2}}{2}\left(\hat{x}_{j_{\alpha}}-\frac{\alpha_{j_{\alpha}} \hat{V}_\alpha(\hat{ q}_\alpha)}{m_{j_{\alpha}}\omega_{j_{\alpha}}^2}  \right)^2],  \nonumber  \\
\label{eqn:inter}
\end{eqnarray}
where $\alpha$ represents the index for PBMs. In the case of BO+Drude SDF from prior research,\cite{PJT25JCP1} $\hat{H}_{A}^{(\alpha)} $ is the harmonic system with  vibrational frequency $\omega_\alpha$, $\hat{U}_{s\alpha}(\hat{q}_s, \hat{q}_{\alpha})= \zeta_s^B \hat{q}_{\alpha}  (V_{\rm LL}^{(s)}  \hat{q}_{s} +V_{\rm SL}^{(s)}\hat{q}_{s}^2 )$ and $\hat{V}_\alpha(\hat{ q}_\alpha)= \hat{ q}_\alpha$.

In this paper, the PBM is interpreted not merely as a device for optimizing the HEOM parameters derived from the MD trajectory, but as a representation of intermolecular modes coupled to a thermal bath. Accordingly, we consider a framework in which the intramolecular modes $s$ are coupled to pseudo-intermolecular modes $\alpha$, characterized by the Ohmic SDF
\begin{eqnarray}
J_{\alpha}(\omega) = \zeta_{\alpha}\,\omega.
  \label{eq:Ohmic}
\end{eqnarray}
Here, $\zeta_{\alpha}$ denotes the coupling strength between the intermolecular.

When optimizing intramolecular vibrational trajectories, the intermolecular mode is practically treated as an additional bath for the intermolecular mode, denoted by $\alpha$ and described as the BO SDF for the intramolecular mode $s$ expressed as\cite{PJT25JCP1,ZTH26JCP3}
\begin{eqnarray}
J_s^\alpha (\omega) =  \frac{m_s \zeta_s^{\rm B}}{2\pi \omega} \frac{J_\alpha^2 (\omega) \omega_\alpha^2 }{(\omega_\alpha^2 - \omega^2)^2 + J_\alpha^2 (\omega)},
  \label{eq:drudePlusBO}
  \end{eqnarray}
where $\omega_{\alpha}$ is the vibrational frequency of the mode $\alpha$, and 
$\zeta_{s}^{\rm B}$ represents the coupling strength between mode $\alpha$ and mode $s$.\cite{GargOnuchic,TT09JPSJ,TT10JCP,DT15JCP} 
Because mode $\alpha$ is introduced solely to optimize vibrational trajectories, its coupling strength $\zeta_{s}^{\rm B}$ tends to be overestimated in spectral calculations. 
Consequently, mode $\alpha$ is treated as a PBM and excluded from IR spectral analyses.

The reduced equation of motion for Eq.~\eqref{eqn:H_total}, excluding $\hat{H}_{\mathrm{inter}}^{\mathrm{PBM}}$, has been derived within the framework of the HEOM formalism. The corresponding classical expression in hierarchical form (CHFPE) is provided in Appendix~\ref{HEOM_Drude}.

\section{sbml4md: Package Overview}
\label{sec:software-architecture}

\begin{figure}[t]
  \centering
  \includegraphics[scale=0.46]{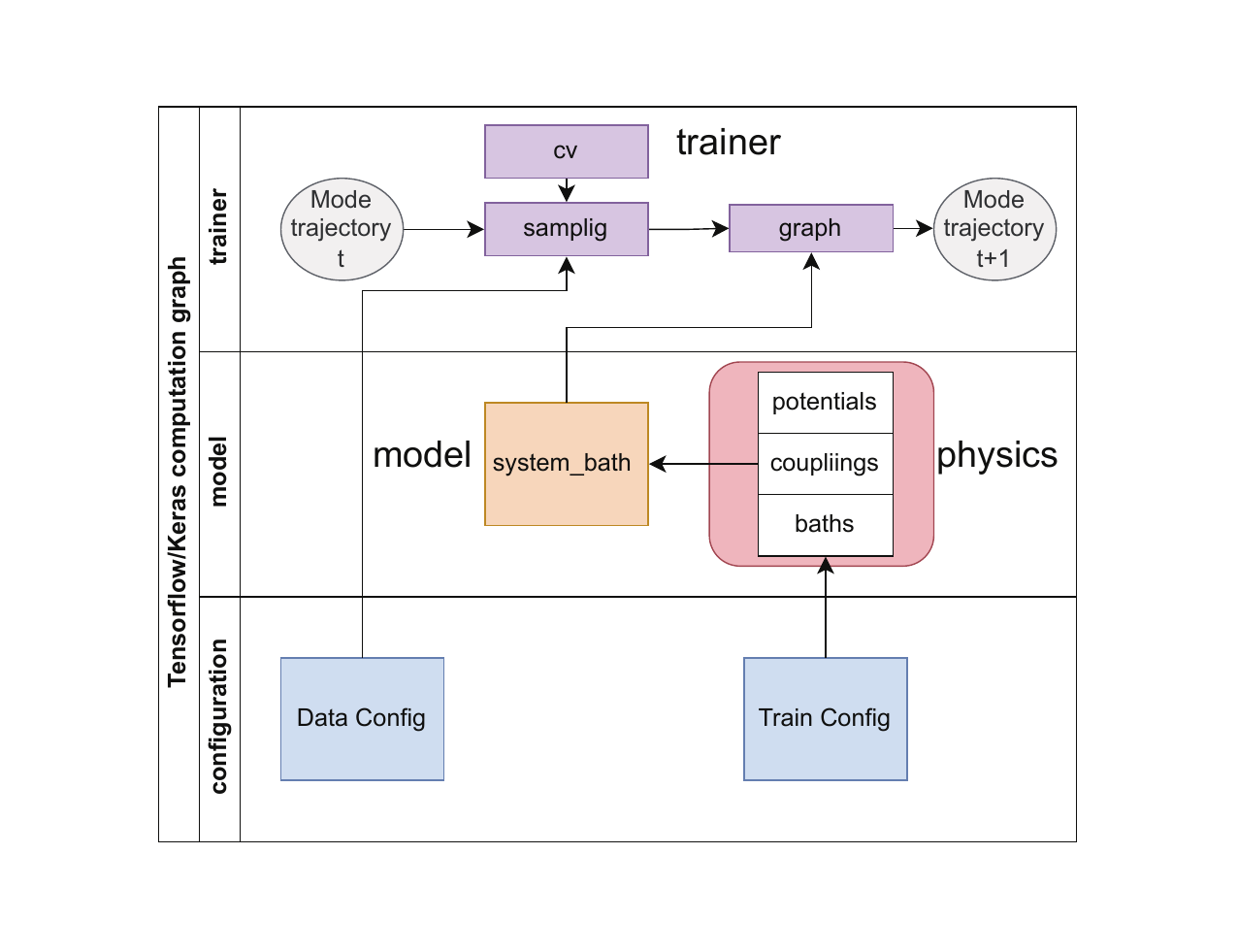}
  \caption{\label{fig:module}Software module architecture.}
\end{figure}

\begin{figure}[t]
  \centering
  \includegraphics[scale=0.77]{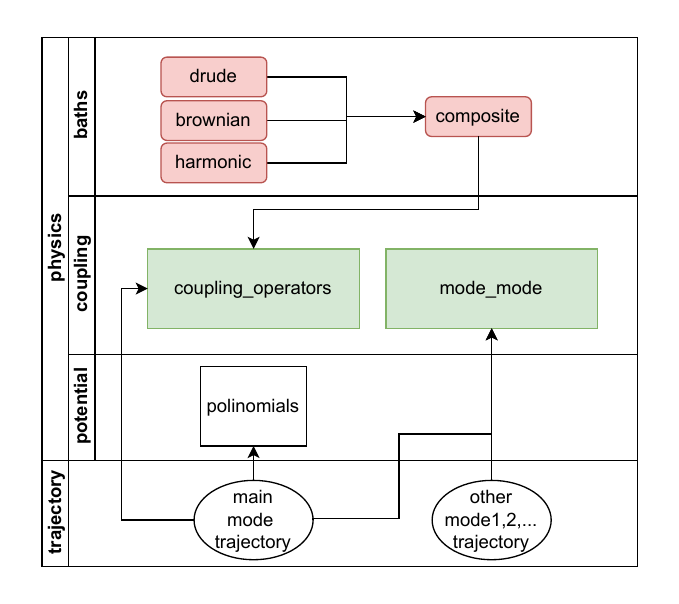}
  \caption{\label{fig:physics}Physics module diagram.}
\end{figure}

Targeting intramolecular vibrational modes, this package constructs and trains S-B models from atomic trajectories obtained through MD, while rigorously maintaining physical consistency. The spectral distribution of intermolecular modes is evaluated from the SDF of the Brownian bath interacting with intramolecular modes. 
Thus, through the present procedure, the anharmonic couplings governing the intramolecular vibrations in Eqs. \eqref{eqn:Potential_s} and \eqref{eqn: Potential ss'}, together with the bath parameters in Eqs. \eqref{eqn: LLSL}, \eqref{eq:drude}, \eqref{eq:Ohmic}, and \eqref{eq:drudePlusBO}, are simultaneously optimized and determined from the MD trajectory.
For details on optimization, see also Refs. \onlinecite{UT20JCTC,PJT25JCP1}.

The package is structured into three top-level modules:
{\it a.} \texttt{physics/}, {\it b.} \texttt{model/}, and {\it c.} \texttt{training/}, 
and theree suplimental modules {\it d.} \texttt{integrators/}, {\it e.} \texttt{config/}, and 
\texttt{execution graph}.  
The top-level components are composed via a lightweight factory into a unified TensorFlow/Keras graph---a computational graph that represents the flow of data through layers and operations---for both training and inference (Fig.~\ref{fig:module}). The design follows object-oriented principles: small, testable primitives (baths, potentials, couplings, integrators) are assembled into a \texttt{SystemBathModel}, which is wrapped by a minimal Keras model exposing standard \texttt{fit}, \texttt{evaluate}, and \texttt{predict} interfaces. This architecture preserves modularity in the scientific core while leveraging Keras's stability and scalability.

\paragraph{\texttt{physics/}---Baths, Potentials, Couplings:} Bath spectral densities are implemented as independent, composable modules: \texttt{Drude}, \texttt{Brownian}, and \texttt{Harmonic} (Fig.~\ref{fig:physics}). ach bath exposes a consistent application programming interface (API, e.g., \texttt{spectral\_weight}) and numerically stable parameterizations for positive variables $\lambda$, $\gamma$, and $\omega_c$. The \texttt{CompositeBath} class allows linear combinations of multiple baths, supporting either global coupling (applied to $\sum J(\omega)$) or per-bath couplings with individual flags (e.g., enabling variance square-linear terms or selecting counter-term modes). Potentials are centered polynomials with trainable harmonic stiffness $k$ and scaled anharmonic terms. Couplings include \texttt{coupling\_operators} (system–bath, VLL/VSL via $V(x)=c_1x+c_2x^2$) and \texttt{mode\_mode} terms (inter-mode LL, SL, LS), with coefficients anchored to the primary mode’s harmonic scale.

\paragraph{\texttt{model/}---S-B Composition and Rollout:} 
The \texttt{SystemBathModel} encapsulates the potential, bath stack, and coupling operators, and exposes the \texttt{rollout\_position} method. Given a batch of input windows $(x_0,\, t_{0:K},\, \text{frequency grid},\, \phi,\, k_BT,\, \text{etc.})$, it evaluates forces from the potential, bath contributions, and optional mode–mode terms, then advances positions using the configured integrator. All components are implemented as \texttt{tf.Module} objects with well-defined \texttt{trainable\_variables}, enabling end-to-end differentiation of $k$, anharmonic scales, $\lambda$, $\gamma$, $\omega_c$, VSL, and mode–mode parameters.

\paragraph{\texttt{training/}---Keras Wrapper, Samplers, and Datasets:} 
The \texttt{RolloutKerasModel} is a thin Keras wrapper that delegates forward passes to \texttt{rollout\_position} and computes a Mean Squared Error (MSE) loss (optionally ignoring the first step for stability). Sampling utilities generate time windows and frequency grids, support random or grid-based initialization, and construct \texttt{tf.data} pipelines for contiguous or expanding time-series splits. Standard callbacks (e.g., Early Stopping, Checkpoint) and optimizers (e.g., Adam) are used without custom training loops. Training artifacts include best weights and a JSON snapshot of learned physical parameters for reproducibility and legacy comparison.

\paragraph{\texttt{integrators/}---Time Propagation:} 
The default integrator uses a position-only kinematic update:
\[
x_{t+\Delta t} = x_t + \tfrac{1}{2}\,a(x_t)\,\Delta t^2,
\]
with optional sub-stepping via \texttt{n\_substeps} and \texttt{recompute\_each\_substep} for improved stability. A Velocity-Verlet (VV) integrator is also available and can be selected for momentum-aware, symplectic propagation.

\paragraph{\texttt{config/}---YAML Schemas and Examples:}
A small factory parses YAML configuration files---YAML denotes a human-readable configuration format used to specify model and training settings---to instantiate the potential, the bath stack (with global or per-bath couplings), the integrator, and the training parameters (e.g., number of steps, stride, batch size, learning rate, patience).
\textbf{Note on units:} all inputs must use mutually consistent units (e.g., time, mass, stiffness, bath rates). Mixing units is a common source of numerical instability or misestimated parameters.

\paragraph{\texttt{execution graph}: Training and Inference:} 
All physics modules are integrated into a unified Keras computation graph. Gradients flow through the entire physics subtree, enabling the training of bath and coupling parameters alongside the potential. Because both inputs and outputs are plain tensors and the datasets are native to \texttt{tf.data}, inference follows the same computational path as training. This design supports batched prediction, cross-validation, and deployment without requiring any code modifications. Tracked variables include $k$, $\lambda$, $\gamma$, $\omega_c$, VSL, and selected mode–mode coefficients.

\section{Training Pipeline and Model I/O Specification}
\label{sec:IOSpecification}
\subsection{Minimal Input Requirements}
The minimal inputs required to run the code are:
(i) MD-derived trajectories for the selected mode coordinates (time series used to construct supervised rollout windows),
(ii) the MD time step and the thermodynamic temperature $T$ (equivalently $k_\mathrm{B}T$),
and (iii) a YAML configuration specifying (a) the chosen physics components (potential form, bath/PBM types) and initial parameter values,
(b) dataset sampling settings (window length \texttt{steps}, stride, batch size, and optional cross-validation (CV) splitting),
and (c) training hyperparameters (learning rate, epochs, early stopping, and loss options such as \texttt{ignore\_first}).
Optional auxiliary channels (e.g., additional mode trajectories for mode-mode coupling terms) may be provided as aligned inputs under \texttt{other:\textit{name}}.
The outputs consist of optimized parameter sets, diagnostic logs, and spectra computed using the resulting effective model.

\subsection{Rollout Model Interface}
We provide a lightweight Keras wrapper, \texttt{RolloutKerasModel}, whose \texttt{call} method directly invokes \texttt{phys.rollout\_position}. The input signature is minimal and fixed:
\[
(\texttt{x0},\,\texttt{dt},\,\texttt{omega},\,\texttt{t\_window},\,\texttt{phi},\,\texttt{kBT},\,\{\texttt{other:\textit{name}}\})
\]
The output is the predicted position trajectory \( y_{\mathrm{pred}} \in \mathbb{R}^{B \times (\text{steps}+1) \times 1} \).  The input/output specification for \texttt{RolloutKerasModel} is summarized in Table~\ref{IO}.

\begin{table*}[!t]
\centering
%\small
\begin{tabular}{llll}
\hline
\textbf{Key} & \textbf{Shape} & \textbf{Dtype} & \textbf{Description / Source} \\
\hline
\texttt{x0} & $[B,1,1]$ & float32 & Initial position at the beginning of the window \\
\texttt{dt} & $[]$ & float32 & Time step (scalar) \\
\texttt{omega} & $[\Omega]$ & float32 & Frequency grid used for driving or sampling \\
\texttt{t\_window} & $[B,\text{steps},1]$ & float32 & Time grid windowed per sequence \\
\texttt{phi} & $[B,1,\Omega]$ & float32 & Phase values for each $\omega$ bin \\
\texttt{kBT} & $[]$ & float32 & Thermal energy (scalar) \\
\texttt{mol\_idx} & $[B]$ & int32 & Molecule or trajectory index (optional) \\
\texttt{other:\textit{name}} & $[B,\text{steps}{+}1,1]$ & float32 & Optional aligned (e.g., mode–mode coupling) \\
\hline
$y$ (target) & $[B,\text{steps}{+}1,1]$ & float32 & Ground-truth trajectory (rollout) \\
$y_{\text{pred}}$ & $[B,\text{steps}{+}1,1]$ & float32 & Predicted trajectory (rollout) \\
\hline
\end{tabular}
\caption{\label{IO}Input/output specification for \texttt{RolloutKerasModel}.}
\end{table*}

\subsection{Dataset Construction and Sampling}
Mini-batches are generated using a sampler configured via \texttt{SamplerConfig}:
\[
\texttt{steps},\;\texttt{stride},\;\texttt{omega\_min},\;\texttt{omega\_max},\;\texttt{omega\_points}.
\]
Each training sample yields a pair \texttt{(inputs, $y$)} matching the I/O specification above. Auxiliary time series, if provided, are automatically included using the naming convention \texttt{other:\textit{name}} and aligned to the same time windows. 
The $\omega$ grid and corresponding phases are generated on the fly as a uniform grid ranging from \texttt{omega\_min} to \texttt{omega\_max}, discretized into \texttt{omega\_points} bins.

\subsection{Configuration-Driven Entry Point}
\begin{lstlisting}[caption={Main training script}, label={lst:main}]
# main.py (abridged)
import yaml
from mab_ml.training.run import run_training
from mab_ml.training.cv import DataCfg, TrainCfg

# Load configuration from YAML
with open("config.yaml") as f:
    cfg = yaml.safe_load(f)

# Instantiate configuration objects
pcfg = cfg["physics"]              # Physical parameters
dcfg = DataCfg(**cfg["data"])      # Data sampling settings
tcfg = TrainCfg(**cfg["train"])    # Training hyperparameters

# Load input arrays
X, dt, others = load_arrays(cfg["data"]["source"])

# Launch training
results = run_training(X, dt, pcfg, dcfg, tcfg, others=others)
\end{lstlisting}
Training is initiated via a compact entry script that reads a YAML configuration file, constructs the necessary objects, and launches the training routine. 

\subsubsection{The core steps}
\begin{enumerate}
  \item Load the YAML configuration.
  \item Extract and instantiate configuration objects for physics, data sampling, and training.
  \item Load input arrays from disk or preprocessing.
  \item Call the training function with all components.
\end{enumerate}
The Python code is presented in List \ref{lst:main}.

\subsubsection{YAML Key Routing Summary}

All scalar values under \texttt{physics} are treated as initial values for trainable parameters. Boolean flags and count-type fields are considered configuration and are not subject to training.

\begin{itemize}
  \item \textbf{\texttt{physics}} $\rightarrow$ \texttt{build\_physics\_and\_model(...)}:
    \begin{itemize}
      \item \texttt{mass} (configuration only), \texttt{kBT}, or the combination of \texttt{temperature\_K} and \texttt{kbt\_unit} (input values; not trainable).
      \item \texttt{potential}: initial values for \texttt{k}, \texttt{x0}, and optional anharmonic terms such as \texttt{cubic}.
      \item \texttt{bath}:
        \begin{itemize}
          \item \texttt{drude}: initial values for \texttt{lambda}, \texttt{gamma}, and optionally \texttt{vsl}.
          \item \texttt{brownian}: initial values for \texttt{lambda}, \texttt{gamma}, \texttt{wc}, and optionally \texttt{vsl}.
          \item \texttt{harmonic}: \texttt{bamp} (trainable if specified), and configuration-only fields \texttt{nomega}, \texttt{wmin}, \texttt{wmax} (used to define the discrete frequency bank).
        \end{itemize}
      \item \texttt{coupling}: initial values for \texttt{g\_linear} (VLL) and \texttt{g\_square} (VSL). Counter-term settings \texttt{use\_counter\_term} and \texttt{counter\_term\_mode} are orthogonal to VLL/VSL and do not toggle VSL. Optional \texttt{couplings} allows per-bath overrides using the same fields.
      \item \texttt{mode\_mode}: for each partner \texttt{name}, initial values for coefficients \texttt{g\_ll}, \texttt{g\_sl}, and \texttt{g\_ls}. Corresponding input channels must be provided as \texttt{other:\textit{name}}.
    \end{itemize}

  \item \textbf{\texttt{data}} $\rightarrow$ \texttt{DataCfg} / sampler (configuration only):
    \begin{itemize}
      \item \texttt{steps} (integration steps per window; output length is \(\texttt{steps}{+}1\)),
            \texttt{stride} (window shift),
            \texttt{batch\_size},
            \texttt{omega\_min}, \texttt{omega\_max}, \texttt{omega\_points} (defines the frequency grid),
            \texttt{steps\_per\_epoch}, \texttt{val\_steps}, and optional \texttt{source} (loader hook).
    \end{itemize}

  \item \textbf{\texttt{train}} $\rightarrow$ \texttt{TrainCfg} (configuration only):
    \texttt{lr}, \texttt{epochs}, \texttt{patience}, \texttt{ignore\_first}, \texttt{seed}, and optional \texttt{ckpt\_path}.

  \item \textbf{\texttt{cv}} (optional; configuration only):
    \texttt{mode} = \texttt{molecule} or \texttt{time}; \texttt{k} (number of folds for molecule-CV) or \texttt{time\_blocks} (explicit ranges \([t_{\min}, t_{\max})\)). Used solely to compute sampler indices.
\end{itemize}
\subsection{Cross-Validation (CV) Modes}

We support two mutually exclusive CV regimes, each with deterministic index computation:

\begin{enumerate}
  \item \textbf{Molecule-CV} (K-fold partitioning along the batch/molecule axis):  
  Let $B$ be the number of molecules or trajectories, and $K$ the number of folds.  
  The set $\{0,\ldots,B{-}1\}$ is divided into $K$ disjoint folds, and training/validation splits are iterated accordingly.  
  During sampling, window start indices are restricted to molecules assigned to the current training fold.

  \item \textbf{Time-CV} (contiguous or expanding windows along the time axis):  
  Let $T$ denote the full trajectory length.  
  We either (a) partition $\{0,\ldots,T{-}1\}$ into contiguous time folds, or (b) use user-defined time blocks.  
  A window start time $t_0$ is considered valid only if the full window $[t_0,\,t_0{+}\texttt{steps}]$ lies entirely within either the training or validation set.
\end{enumerate}

\noindent\textit{Schematic illustration:}
\[
\underbrace{\boxed{\text{mol train}}\;\boxed{\text{mol val}}}_{\text{Molecule-CV}}
\qquad
\underbrace{\boxed{\text{past train}}\;\boxed{\text{val}}}_{\text{Time-CV (expanding)}}
\]

Both regimes preserve the exact rollout input/output structure and do not modify the model architecture.  
They only affect how the sampler selects window indices for each training batch.

\subsection{Trainer: Optimizer, Learning Rate, and Loss}
\label{sec:trainer}

\paragraph{Optimizer and learning rate policy:} 
Training is performed using the Adam optimizer with an initial learning rate \(\eta_0\). By default, the learning rate remains constant:
\[
\eta(t) = \eta_0,
\]
but simple decay schemes are also supported, such as exponential decay:
\[
\eta(t) = \eta_0 \exp(-t/\tau),
\]
where \(\tau\) is the decay time constant. Early stopping based on validation loss and checkpointing of the best-performing model are used to improve generalization and reproducibility.

\paragraph{Rollout loss: MSE with optional first-step exclusion:}

Each training sample provides a rollout window of length \(S{+}1\), with time points \(t_s = s\,\Delta t\) for \(s = 0, \dots, S\). Let \(y_{b,s} \in \mathbb{R}\) denote the ground-truth trajectory for batch index \(b\) and time index \(s\), and \(\hat{y}_{b,s}\) the corresponding model prediction. To emphasize dynamic evolution rather than trivial matching of initial conditions, the first step (\(s = 0\)) can optionally be excluded from the loss. Define the time index set:
\[
\mathcal{S} =
\begin{cases}
\{1,2,\dots,S\}, & \text{if \texttt{ignore\_first} = True},\\[2pt]
\{0,1,\dots,S\}, & \text{otherwise}.
\end{cases}
\]
The MSE over batch and time is then:
\[
\mathcal{L}_{\mathrm{MSE}}
=
\frac{1}{B\,|\mathcal{S}|}
\sum_{b=1}^{B}
\sum_{s\in\mathcal{S}}
\bigl\|\hat y_{b,s} - y_{b,s}\bigr\|_2^2.
\]
In practice, this corresponds to slicing away the first time index when \texttt{ignore\_first} is enabled, followed by reduction over \((b, s)\).

\paragraph{Windowing rationale:} 
Training on short windows ($S{+}1$ steps) stabilizes optimization by
providing many overlapping supervision segments sampled from long trajectories
(with user–controlled stride and start indices).
Ignoring $s{=}0$ prevents the loss from being dominated by the enforced initial
condition and focuses the objective on the one–step–ahead (and longer–horizon)
evolution error.

\paragraph{Optional extensions (time weights and multi-task weighting):} 
If desired, per-time weights \(\{w_s\}_{s \in \mathcal{S}}\) can be used to emphasize specific horizons:
\[
\mathcal{L}_{\mathrm{MSE},w}
=
\frac{1}{B\,\sum_{s\in\mathcal{S}} w_s}
\sum_{b=1}^{B}
\sum_{s\in\mathcal{S}}
w_s\,\bigl\|\hat y_{b,s} - y_{b,s}\bigr\|_2^2.
\]
For multiple targets or output heads indexed by \(k = 1, \dots, K\), a weighted sum is used with scalar hyperparameters \(\{\alpha_k\}\):
\[
\mathcal{L}_{\mathrm{total}}
=
\sum_{k=1}^{K} \alpha_k\,\mathcal{L}_k,
\qquad
\alpha_k > 0,\quad \sum_{k=1}^{K}\alpha_k = 1,
\]
where each \(\mathcal{L}_k\) follows the same rollout loss definition, optionally excluding the first step.

\paragraph{Default settings (this work).}
We use Adam with constant learning rate \(\eta_0\), early stopping on validation loss, and rollout length \(S{+}1\) with \texttt{ignore\_first} = True. User-exposed hyperparameters include:
\begin{itemize}
  \item Initial learning rate \(\eta_0\)
  \item (Optional) decay time \(\tau\)
  \item Rollout length \(S\) and window stride
  \item (If enabled) time weights \(\{w_s\}\) and task weights \(\{\alpha_k\}\)
\end{itemize}

\subsection*{Limitations and Intended Use}

This package models a single one-dimensional primary mode evolving in a polynomial potential---harmonic with optional cubic or higher-order anharmonic terms---driven by an arbitrary collection of baths (Drude, underdamped Brownian, and/or harmonic oscillator components). Each bath contributes its own spectral density and may carry its own coupling operator (LL, SL, or both). Optional mode–mode couplings allow other coordinates to enter as exogenous series.

By default, the system uses a position-only kinematic updater (no explicit momenta, non-symplectic), so numerical stability depends on choosing a sufficiently small time step $\Delta t$ relative to the stiffest curvature and damping.velocity-Verlet integrator is available when momentum resolution is needed, though it still assumes a single coordinate system.

Bath driving is controlled by the SDF, with Drude and BO terms contributing at the kernel level. Counter-term choices (``legacy'' vs.\ ``exact'') modify the effective stiffness experienced by the mode via
\[
k_{\mathrm{eff}} = k - \sum_b \Lambda_b.
\]
This model is intended for use in regimes where:
\begin{enumerate}
  \item Dynamics occur near a single potential well, and a 1D coordinate captures the dominant curvature.
  \item Anharmonicity is weak to moderate.
  \item Rollouts are short to moderate in length, and colored-noise baths serve as compact descriptors of environmental memory.
\end{enumerate}

Outside this regime---e.g., strongly coupled multi-dimensional dynamics, double-well transport, or dynamics requiring momentum---this model is descriptive at best and may mis-specify force balance.

\paragraph{Identifiability in short/noisy windows:} 
Training uses short windows (length \texttt{steps}, optionally subsampled) with random start indices, a discrete \(\omega\) grid, and a phase field \(\phi\). Short or noisy windows can make parameters such as \((\lambda, \gamma, \omega_c)\) and coupling gains (e.g., VLL vs.\ VSL) empirically similar in effect.

The rollout MSE loss,\[
\mathcal{L} = \frac{1}{S} \sum_{t = t_0 + 1}^{t_0 + S} \left\| \hat{x}_t - x_t \right\|^2,
\]
optionally ignores the first step to reduce sensitivity to initial transients. While this improves optimization, it further weakens parameter constraints when \(S\) is small.

To ensure numerical robustness, parameters are mapped through positivity-enforcing transforms (e.g., softplus for rates). These transforms help stability but can mask weak identifiability or bias estimates under heavy noise. Brownian/damped-oscillator parameterizations often include practical constraints (e.g., \(\gamma\) tied to or bounded by \(\omega_c\)), which regularize fits but also restrict admissible solutions.

\paragraph{Unit sensitivity:} 
All parameters must be unit-consistent. Common failure modes include:
\begin{itemize}
  \item Mixing stiffness \(k\) specified as a target wavenumber (cm\(^{-1}\)) with \(k\) interpreted as mass–frequency stiffness (amu/ps\(^2\)).
  \item Providing \(\lambda, \gamma, \omega_c\) in inconsistent time units.
\end{itemize}

Thermal energy \(k_\mathrm{B}T\) is resolved from user input (e.g., K or cm\(^{-1}\)) into the internal unit system. Any mismatch affects the noise scale and apparent stiffness. Since acceleration is derived from force divided by mass, an incorrect mass or scaling factor (\texttt{pos\_scale}) rescales both the learned \(k\) and bath amplitudes.

\paragraph{Dependence on mode extraction.}
The primary series is typically derived from molecular trajectories after:
\begin{itemize}
  \item Center-of-mass removal
  \item Rotational alignment
  \item Projection onto a chosen mode
\end{itemize}
Optional ``other:'' channels feed mode–mode coupling terms.

Errors in these preprocessing steps---projection formulas, centering, unit conversion, windowing/stride---propagate \emph{directly} into fitted parameters. Trained descriptors generalize best when the same preprocessing, \(\omega\) grid, \(\Delta t\), and scaling are preserved. Changing any of these (e.g., sampling rate or normal-mode basis) generally requires re-identification.

\subsection*{Availability and Reproducibility}

The pipeline is configuration-driven. A YAML file specifies:
\begin{itemize}
  \item \textbf{Physics:} potential coefficients; list of baths with per-bath couplings; counter-term and square-term flags
  \item \textbf{Data:} window length (\texttt{steps}), stride, discrete \(\omega\) grid and phase handling, batch size, seeds
  \item \textbf{Training:} optimizer and learning rate policy, loss with \texttt{ignore\_first}, early-stopping criteria, number of epochs
\end{itemize}

The entry point parses the YAML, constructs data/physics/training configurations, builds the computational graph, and launches training. The sampler is deterministic given its seed and emits the same windows, \(\omega\) grid, and phase realizations.

Cross-validation supports at least two regimes:
\begin{itemize}
  \item Molecule-wise (split by entity)
  \item Time-wise (contiguous blocks)
\end{itemize}
Indices are computed explicitly and stored.

For exact replay, the code saves:
\begin{enumerate}
  \item The full YAML configuration
  \item Machine-readable parameter snapshots (trainable potential/bath/coupling values and training hyperparameters)
  \item Model weights
  \item Epoch-wise metrics and the selection rule for the ``best'' checkpoint
\end{enumerate}

To ensure portability, users should also archive:
\begin{itemize}
  \item Raw or preprocessed trajectories
  \item Unit declarations (how \(k\), \(\lambda, \gamma, \omega_c\), and \(k_\mathrm{B}T\) were specified)
  \item Mass used for the primary mode
  \item \(\omega\) grid and \texttt{pos\_scale}
  \item Sampler seeds
  \item Counter-term and square-term settings
\end{itemize}

With these artifacts, another user can reproduce the full experiment---data windows, training curves, and selected parameters---under the stated 1D-mode assumptions and unit conventions.

\section{Demonstration of Intramolecular Vibrational Modes in Water}
\label{sec:Demo}

We demonstrate the capabilities of the present software by calculating the linear absorption spectrum and the 2D correlation spectrum of intramolecular vibrations in liquid water. In this setting, the solute $s$ corresponds to the intramolecular vibrational modes of a water molecule, whereas the bath (solvent) represents the surrounding water environment. We demonstrate the capabilities of the present software by calculating the linear absorption spectrum and the two‑dimensional correlation spectrum of intramolecular vibrations in liquid water. 
To represent the three primary intramolecular modes of water, we define
 $\bm{q}=(q_1, q_{1'}, q_2)$ for the (1) anti-symmetric stretch, ($1'$) symmetric stretch, and (2) bending modes, respectively.\cite{HT25JCP1,HT26JCP1} 

To accelerate the ML optimization, we additionally introduced PBMs with $\alpha = 1\text{--}3$ coupled to each intramolecular mode $s$. Although explicit intermolecular vibrational modes are not included in the present study, analysis of these auxiliary PBMs indicates that they can also assist in characterizing such intermolecular motions.

\subsection{Training Dataset Derived from MD Simulations}

As a demonstration, we optimize the parameters of the MAB-Drude model based on molecular dynamics (MD) trajectories of water and validate the method by calculating the spectrum. Because we model the fast intermolecular modes, which arise from short-range intermolecular interactions, it is not necessary to carry out large-scale simulations with many molecules.\cite{JIT16CP} Thus, the trajectories were obtained from simulations of 392 water molecules confined in a cubic box of 2.3~nm per side under periodic boundary conditions. MD simulations were performed using GROMACS\cite{GROMACS2025Manual} with (a) the flexible SPC/E water model\cite{ABRAHAM201519} and (b) the Ferguson potential model.\cite{1995JCoCh..16..501F} For both models, the system was equilibrated at 300~K, and trajectories were generated with a time step of 0.1~fs over a total simulation time of 50~ps.

Note that our single‑molecule–trajectory‑based approach does not sample the full phase space because the ML workflow uses short rollout windows and does not assume ergodic, long‑time behavior. Yet the effective bath parameters arise from slow intramolecular couplings with long environmental correlation times, which cannot be captured from short‑time training. Consequently, incorporating inhomogeneous broadening effects---represented in the MAB model by non‑Markovian system–bath interactions and finite noise correlation times—is challenging.

Without changing the trajectory length, we then found that these inhomogeneous effects can be captured by increasing the initial deviation of the sampled molecule and by further reducing the early‑stopping distance.

To account for the influence of intermolecular modes on intramolecular vibrations, our software is designed to incorporate an arbitrary number of PBMs. For liquid water, three representative intermolecular modes were intend to introduced: (3) hindered rotation (HR), with 600 cm$^{-1}$ (4) high-frequency translational (HT) motion, with 200 cm$^{-1}$, and (5) low-frequency translational (LT) motion, with 60 cm$^{-1}$.\cite{HT25JCP2} It should be noted, however, that the optimized PBMs do not necessarily correspond to these specific modes.

During the optimization process, since prior studies had shown that LL interactions exert only a minor effect on intramolecular vibrations,\cite{PJT25JCP1} we optimized the parameters of them by considering only SL interactions.

For ML, model training was performed using Python 3.9.18 in conjunction with  TensorFlow 2.15 and CUDA 12.2. 
All computations were executed on a system equipped with an Intel Core i9-13900H CPU and an NVIDIA GeForce RTX 4070 GPU. On our hardware, one fold for a single mode takes approximately 2–4 hours, so all three modes across four folds take approximately 24 to 48 hours.

\subsection{Optimized parameter values for the HEOM-Drude model}

The resulting trajectories were transformed into normal mode coordinates corresponding to intramolecular vibrations. We then trained models for the Drude SDF with three intermolecular pseudo modes represented as BO SDF.
For each training and testing split, early stopping was applied with a patience threshold of 300 epochs. Optimization was terminated when the test loss did not improve for 300 consecutive epochs.  For details on ML, refer to Ref. \onlinecite{PJT25JCP1}.

The dipole function is defined as 
\begin{eqnarray}
\label{eq:dipole}
{\mu}(\bm{q})=\sum_{s} (\mu_s q_s +\mu_{ss} q_s^2)+ \sum_{s\ne s'} \mu_{s{s'}} q_s q_{s'},
\end{eqnarray}
where $\mu_s$ and $\mu_{ss'}$ denote the linear and nonlinear components of the dipole moment. In this paper, the values for each polarization were set based on the values in Ref \cite{IT16JCP}.

To facilitate comparison with previous studies using collective modes\cite{IT16JCP,TT23JCP1,TT23JCP2} %and to support 2D spectrum calculations with our quantum simulation code, 
we adopted the formatting conventions of Refs.~\onlinecite{HT25JCP1,HT25JCP2}. The scaled parameters are defined as $\tilde{\zeta_s} \equiv \zeta_s (\omega_0/\omega_s)^2$, with $\omega_0 = 4000\,[\mathrm{cm}^{-1}]$. Mode–mode coupling terms are normalized as
 $\tilde g_{s^3}=\overline{g}_{s^3}(\omega_0/\omega_s)^3$, $\tilde g_{s's}=\overline{ g}_{s's}(\omega_0/\omega_s)(\omega_0/\omega{s'})$, $\tilde g_{s^2 s'}=\overline{g}_{s^2 s'}(\omega_0/\omega_s)^2(\omega_0/\omega{s'})$, and $\tilde g_{s s'^2}=\overline{g}_{s s'^2}(\omega_0/\omega_s)(\omega_0/\omega{s'})^2$.

The results for the optimized MAB-Drude parameters, PBM bath parameters, and dipole elements used in the subsequent IR calculations for MD trajectories generated with the flexible SPC/E water model and the Ferguson potential model are summarized in Appendix~\ref{SPCE_PARA}.

\subsection{Classical Hierarchical Fokker--Planck Equation for three-mode MAB model (CHFPE-2DVS)}

Using a machine‑learned MAB model for the three intramolecular modes of water, we assess its accuracy by computing the 1D and 2D correlation spectra through the recently developed CHFPE for 2DVS that is publicly available.\cite{HT25JCP1,HT25JCP2} The CHFPE is provided in Appendix~\ref{HEOM_Drude}.

Although the vibrational modes are treated classically, the underlying MD trajectories were also generated using classical dynamics. In this sense, the consistency between the IR spectra and the MD results may be more direct than in quantum treatments.

\subsubsection{Linear absorption spectra}
\label{LAS}

The linear absorption spectrum is calculated from the first-order response function, expressed as\cite{IT16JCP,TT23JCP1,HT25JCP1}
\begin{eqnarray}
R^{(1)}(t) = \frac{i}{\hbar} \mathrm{Tr} \left\{ \hat{\mu} \hat{\mathcal{G}}(t) \hat{\mu}^{\times} \hat{\rho}^{\mathrm{eq}} \right\}.
\label{R_1t}
\end{eqnarray}
Here, $\hat{\mathcal{G}}(t)$ denotes the Green's function associated with Eq.~\eqref{eq:clHEOM}, and $\hat{\rho}^{\mathrm{eq}}$ is the equilibrium density operator or phase-space distribution obtained from the steady-state solution of the HEOM. 

To evaluate $R^{(1)}(t)$,  the HEOM is propagated from the initial condition $\hat{\mu}^{\times} \hat{\rho}^{\mathrm{eq}}$ at $t = 0$. The solution at time $t$, denoted by $\hat{\rho}'(t)$, yields the response function via the expectation value:
 $R^{(1)}(t) = i {\rm tr}\{ \hat\mu \hat \rho'(t) \}/\hbar $.\cite{T06JPSJ,IIT15JCP,IT16JCP}

The linear absorption spectrum is then obtained by Fourier transforming the first-order response:
\begin{equation}
\alpha(\omega) = \omega\,\Im\!\int_{0}^{\infty}\!dt\,e^{i\omega t}\,R^{(1)}(t).
\end{equation}

The optimized parameters for the flexible SPC/E model the Ferguson model are listed in Appendix \ref{SPCE_PARA}. We note that the inclusion of intermolecular vibrations---whether treated as PBMs or not---has a negligible effect on the optimized parameters associated with intramolecular vibrations.\cite{PJT25JCP1}.
It is worth noting that the ML approach enables the evaluation of both linear and nonlinear dipolar elements. However, since their difference is negligible in the linear absorption spectrum, normalized values were employed.

\begin{figure}[!t]
  \centering
  \includegraphics[scale=0.37]{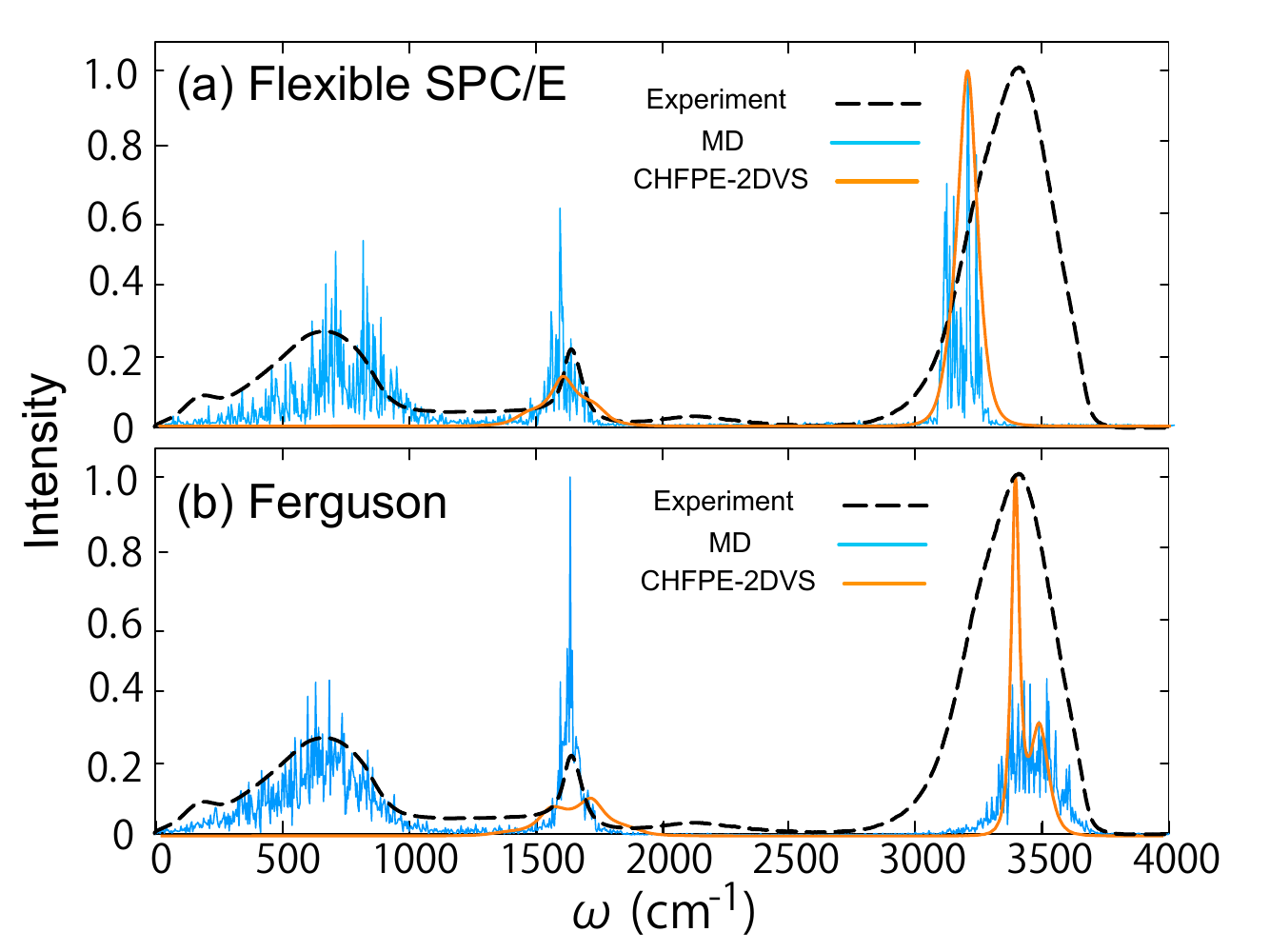}
    \caption{\label{fgr:SPCE_ir_vll}Infrared absorption spectra evaluated using (a) the flexible SPC/E potential and (b) the Ferguson potential for water are shown. The spectra were calculated with CHFPE using MAB parameters trained on MD trajectories generated with the flexible SPC/E and Ferguson potentials. Both the stretching and bending vibrational modes are clearly observed. For comparison, each panel also 
includes results from MD simulations (blue lines) and experimental data 
(black dashed curves).\cite{IRexp2011} The experimental spectrum is reproduced with permission 
from Author, J. Mol. Struct. 1004, 146 (2011). Copyright (2011) Elsevier.
} 
\end{figure}

To evaluate Eq. \eqref{R_1t}, we integrate the CHFPE in Eq. \eqref{eq:clHEOM} for three-intramolecular modes.\cite{HT25JCP1,HT25JCP2} Figure \ref{fgr:SPCE_ir_vll} presents the calculated spectra for the bending and stretching modes using the (a) flexible SPC/E potential and (b) Ferguson potential. For comparison, spectra derived directly from the MD trajectories (light blue lines) and experimental measurements (black dashed curves) are also plotted.

It is worth emphasizing that while the (a) flexible SPC/E model allows for intramolecular vibrational motion, its treatment of anharmonicity is limited due to the use of relatively simple potential functions. Consequently, in the IR spectra computed directly from MD trajectories, the flexible SPC/E model yields two distinct peaks corresponding to the symmetric and anti-symmetric stretching modes. In contrast, the Ferguson model, which incorporates a more sophisticated treatment of anharmonicity and intermolecular coupling, produces broadened peaks in the MD‑derived spectra, making the two stretching modes difficult to distinguish.

The peak widths of the stretch modes obtained from the CHFPE-2DVS calculations using the MAB model, evaluated through the ML approach in Fig. \ref{fgr:SPCE_ir_vll}, agree well with those obtained directly from the molecular dynamics simulations (light-blue lines). The peak broadening is smaller than the experimental values, which is likely because classical calculations do not include broadening arising from zero-point vibrational motion. The MD stretching results for the Ferguson potential show symmetric and anti-symmetric stretching splitting apart, and, reflecting this structure,
the CHFPE-2DVS results also split. The results for the Ferguson potential
also show bending splitting, which is likely due to the S-B coupling being overestimated when optimizing the HEOM parameters to match the strong bending peak observed in the MD results.

The MD-derived peak positions of flexible SPC/E result lie below the experimental values, reflecting limitations inherent in the classical potential functions used in MD simulations. The optimized peak positions follow a similar trend.  These deviation likely due to the low anharmonicity of the optimized parameters. The MD-derived peak positions for the flexible SPC/E model lie below the experimental values, reflecting limitations inherent in the classical potential functions employed in MD simulations. The optimized peak positions exhibit a similar trend. These deviations are likely attributable to the low anharmonicity of the optimized parameters.
In contrast, the peak positions obtained from the Ferguson potential agree remarkably well with the experimental values, presumably because the potential parameters were fitted so as to reproduce classical peak positions. However, such peak alignment through parameter fitting is not an appropriate strategy for our purposes, as our objective is to perform quantum simulations using the MAB model developed in this work. When the same model is used in quantum simulations, the peak positions are expected to undergo a red shift.\cite{HT11JPCB,ST11JPCA}
For the flexible SPC/E model, the deviation from the experimental peak positions is expected to be even larger. Therefore, these potentials are not suitable for our modeling.

\begin{figure}[!t]
  \centering
  \includegraphics[scale=0.3]{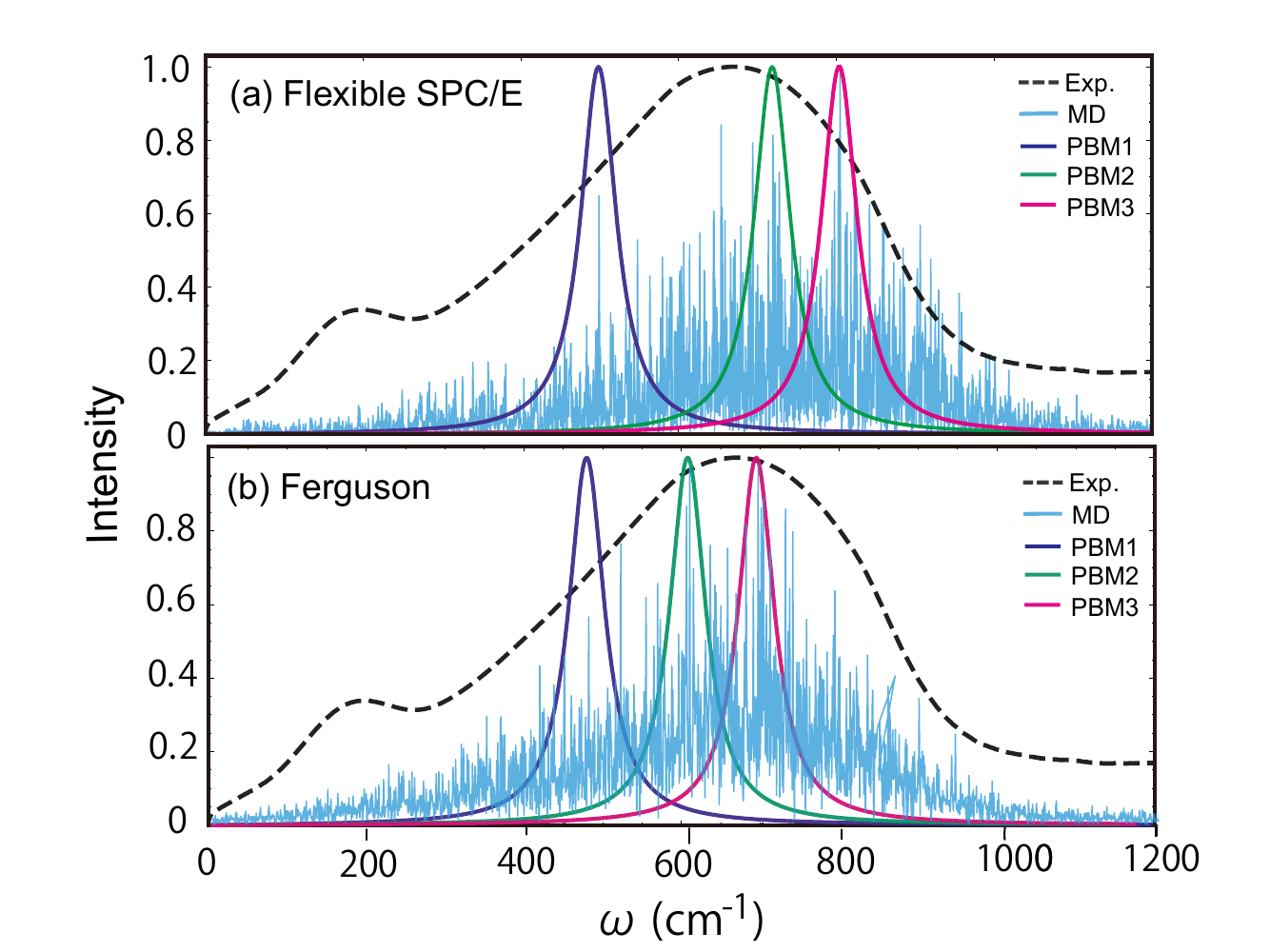}
  \caption{\label{fgr:spce_ir}
  PBM spectra in the low-frequency region for the (a) flexible SPC/E potential and (b) Ferguson potential.
  Here, PBM1 (purple), PBM2 (green), and PBM3 (magenta) are plotted using the fitted PBM parameters listed in
  Table~\ref{tab:spce_drude_3bo_compact_products}.
  For comparison, each panel also includes results from MD simulations (blue lines) and experimental data (black dashed curves).\cite{IRexp2011} The experimental spectrum is reproduced with permission 
from Author, J. Mol. Struct. 1004, 146 (2011). Copyright (2011) Elsevier.
  }
\end{figure}

Figure \ref{fgr:spce_ir} presents spectral profiles constructed to examine whether the PBM feature arises from intermolecular vibrations. Three representative modes were introduced to mimic typical intermolecular vibrational motions; however, the optimized peaks coincided with those observed in MD simulations. This correspondence likely reflects the assumption of an Ohmic thermal bath for PBM, which favors sharp spectral features. Employing a Drude-type bath may instead produce broader peaks spanning the entire region.

We now explore the characteristics of PBMs introduced for effective trajectories optimization. Although the intensity of the PBM corresponding to the absorption spectrum cannot be evaluated, its frequency distribution is recognized as being described by the distribution in Eq. \eqref{eq:drudePlusBO} with Eq. \eqref{eq:Ohmic}.  

We then found that, although the individual frequencies do not directly correspond to the representative intermolecular modes,\cite{HT25JCP2} they generally align with the spectral features extracted from the MD trajectories. These peaks may originate from localized vibrational modes induced by intermolecular cage effects.

The failure of PBM to accurately reproduce the distribution of intermolecular vibrations was likely due to the assumption of an Ohmic SDF (Eq. \eqref{eq:Ohmic}.) for the bath coupled to the PBM, which does not account for inhomogeneous broadening effects. Within this ML framework, the Drude SDF [\eqref{eq:drude}] can be introduced as an alternative to the Ohmic mode. Further exploration is planned regarding the role of PBM.

\subsubsection{2D correlation IR Spectra}

\begin{figure}[!t]
  \centering
  \includegraphics[scale=0.5]{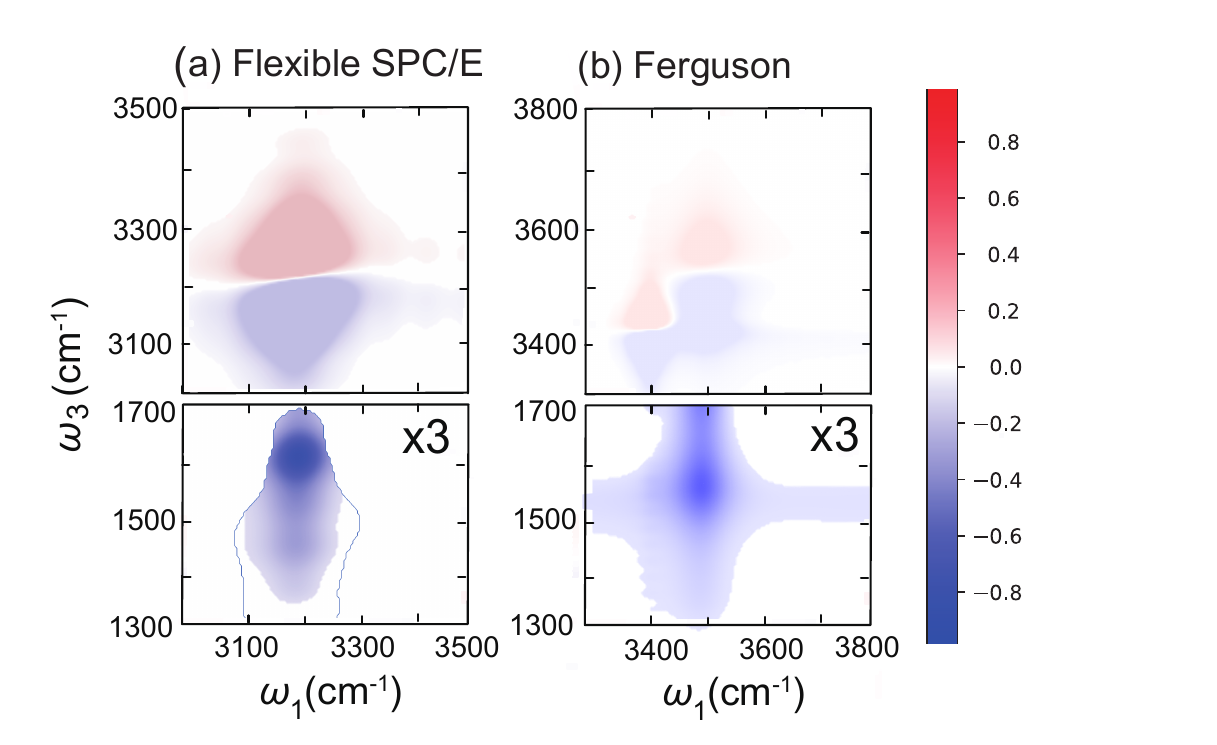}
    \caption{\label{fgr:2Dst-bnd}2D correlation IR spectra for the stretching and stretching$\rightarrow$
bending motions were obtained using optimized parameter values from (a) the flexible SPC/E model and (b) the Ferguson model. Because the peak intensities in the lower panels are weaker than those in the upper panels, the contour
interval was tripled to enhance visibility.} 
\end{figure}

\begin{figure}[!t]
  \centering
  \includegraphics[scale=0.5]{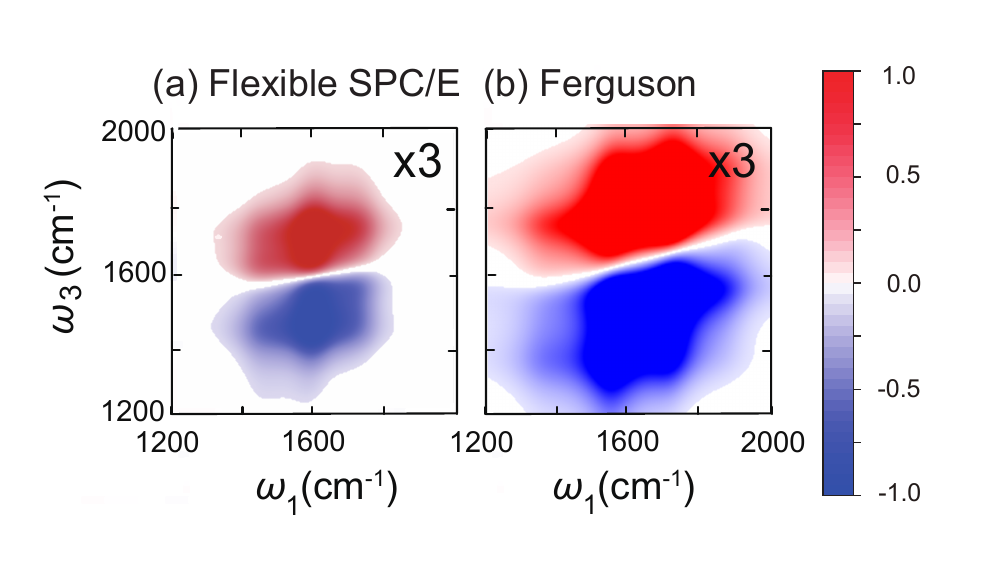}
    \caption{\label{fgr:2Dbnd}Results for the same calculations as in Fig.~\ref{fgr:2Dst-bnd}, but 
for the bending mode. Because the peak intensity 
is weaker than in the upper panel of Fig.~\ref{fgr:2Dst-bnd}, the contour
interval was tripled for clarity.} 
\end{figure}

The three-body response function of the dipole moment is given by\cite{T06JPSJ}
\begin{eqnarray}
\label{BO:2D}
R(t_3,t_2,t_1) =  \mathrm{Tr} \left\{ \hat {\mu}\, \mathcal{G}(t_3)\, {\hat \mu}^{\times}
\mathcal{G}(t_2)  {\hat \mu}^{\times}\mathcal{G}(t_1)\, {\hat \mu}^{\times}
\hat{\rho}^{\mathrm{eq}} \right\}, \nonumber \\
\end{eqnarray}
The 2D correlation IR spectrum is obtained by combining, with equal weights, the rephasing
$R_{R}(t_3,t_2,t_1)$ and non-rephasing $R_{NR}(t_3,t_2,t_1)$ contributions.\cite{2DCrrJonas2001,2DCrrGe2002,2DCrrTokmakoff2003}
A widely used definition of the 2D correlation signal is\cite{2DCrrGe2002,2DCrrTokmakoff2003,TI09ACR}
\begin{align}
&I_{\rm C}(\omega_3,t_2,\omega_1) \nonumber \\
&= {\rm Im}\left\{
\int_0^\infty dt_1 \int_0^\infty dt_3\,
e^{-i\omega_1 t_1} e^{+i\omega_3 t_3}
R_{\rm R}(t_3,t_2,t_1)
\right\} \nonumber \\
&\quad + {\rm Im}\left\{
\int_0^\infty dt_1 \int_0^\infty dt_3\,
e^{+i\omega_1 t_1} e^{+i\omega_3 t_3}
R_{\rm NR}(t_3,t_2,t_1)
\right\}. \label{2DCorr}
\end{align}

Although, in principle, $R_{\rm R}$ and $R_{\rm NR}$ can be separated by selecting specific Liouville pathways in an energy-level representation, such a decomposition becomes impractical when the response functions are evaluated in phase space.  
To circumvent this difficulty, we suppress the unwanted rephasing contribution by applying a Fourier transform with respect to $t_2$ to $I_{\rm C}(\omega_3,t_2,\omega_1)$, while setting
$R_{\rm R}(t_3,t_2,t_1)=R_{\rm NR}(t_3,t_2,t_1)=R(t_3,t_2,t_1)$.
This procedure removes the oscillatory component with frequency $2\nu_s$, where $\nu_s$ is the frequency of the target mode $s$.\cite{TI09ACR,HT08JCP,YagasakiSaitoJCP20082DIR,TT23JCP2}

Fig.~\ref{fgr:2Dst-bnd} shows the 2D correlation IR spectra for the stretching motion and the stretching$\rightarrow$bending motion obtained from the classical calculations under the same physical conditions as in the 1D case. The calculations were performed using a three-mode model consisting of (1) the OH stretching mode, (1$'$) the OH anti-stretching mode, and (2) the HOH bending mode for (a) the flexible SPC/E model and (b) the Ferguson model.

In the present classical calculation, the two separated positive (red) and negative (blue) peaks originate from anharmonic motion near the bottom of the potential. In contrast, in the quantum case, the red peak corresponds to the $0\!-\!1\!-\!0$ transition and the blue peak to the $0\!-\!1\!-\!2$ transition, where 0, 1, and 2 denote the energy eigenstates of the vibrational mode.\cite{ST11JPCA,HT25JCP1}  The stretch-mode results for the Ferguson model show two paired peaks, positive and negative, reflecting the symmetric and anti-symmetric splitting observed in the linear absorption spectrum.

An inclination of the node line appears when the intramolecular modes interact with the environment in a non-Markovian manner. In contrast, no such feature is observed in the cross-peak associated with the stretch--bend coupling. These observations are consistent with the classical three-mode analysis previously performed for the intramolecular modes of water.

The bending peaks shown in Fig.~\ref{fgr:2Dbnd} exhibit a trend consistent with earlier analyses of 2DIR measurements, although the broadening is considerably larger. In particular, for the Ferguson model, the peak splitting observed in the linear absorption spectrum leads to a substantial increase in the bend-peak width. The observed splitting may result from modeling the strong Lorentz‑like bending peak in the Ferguson model exclusively through the SL interaction. Since energy relaxation plays a larger role in low‑frequency bending modes, incorporating LL interactions may be required.

This broadening is likely attributable to limitations in the molecular dynamics description of the bending modes. As demonstrated here, the optimized MAB parameters are highly sensitive to the
quality of the underlying MD potential.

Building on the linear absorption results, we emphasize that neither the flexible SPC/E model nor the Ferguson model provides a physically adequate basis for constructing a quantum-oriented MAB parameter set. Consequently, the present work is limited to presenting representative snapshots of the 2D correlation spectra. A full analysis will be pursued once MD trajectories specifically tailored for quantum simulations become available, as these are essential for advancing the next stage of this research.

\section{Conclusion}
\label{sec:conclusion}

This study represents an initial step toward establishing an open quantum dynamic framework for molecular liquids, grounded in MD simulation and aimed at computing and interpreting nonlinear vibrational spectra. This technique can be applied to probing solute molecules in arbitrary solutions as well as reaction centers within biomolecules.

As a concrete demonstration, we focused on the ultrafast nonlinear vibrational spectroscopy of water.
Although individual water molecules are structurally simple, the collective dynamics they exhibit are strikingly rich and central to life. These emergent phenomena are governed by fundamental physical laws, and ML offers a powerful means to uncover the underlying mechanisms and deepen our physical understanding. While the present ML-based framework has proven useful for analysis, it has not yet succeeded in reproducing experimentally observed spectra with quantitative accuracy.

To rigorously validate this approach, it will be essential to develop a dedicated simulation platform capable of computing high-fidelity 2D spectra within a quantum mechanically rigorous framework.\cite{HT26JCP1} Equally important is the quality of the training data: improvements in the underlying MD force fields and quantum dynamical simulations are indispensable. Even in the era of ML, there are no shortcuts---only through sustained, systematic refinement can one achieve reliable and physically meaningful results. 

Although the full realization of this vision remains a work in progress, the present study provides a prototype implementation and a foundation upon which future developments can build. A more detailed analysis will be presented in a subsequent paper using a more descriptive MD potential.

\section*{Acknowledgments}
Y. T. would like to express his sincere gratitude to Professor Shinji Saito for his insightful suggestions and valuable advice on various issues related to water simulations.
Y. T. was supported by JST (Grant No. CREST 1002405000170). K. P. acknowledges a fellowship supported by JST SPRING, the establishment of university fellowships toward the creation of science technology innovation (Grant No.~JPMJSP2110). 

\section*{Supplementary Material}

The supplementary material includes the \texttt{sbml4md} software, together with demonstration codes for HEOM modeling using molecular dynamics trajectories. Detailed instructions are provided in the \texttt{ReadMe.tex} file.

\subsection*{Conflict of Interest}
The authors have no conflicts to disclose.

\section*{Data availability}
The data that support the findings of this study are available from the corresponding author upon reasonable request.

\appendix

\begin{table*}[!htb]
\caption{\label{tab:spce_drude_3bo_bath}
Optimized $s$th intramolecular mode parameters of the MAB model trained from flexible SPC/E and Ferguson potentials incorporating the Drude SDF with SL interaction and 3 Brownian Oscillator with LL + SL interaction are presented for the vibrational modes: 
(1) anti-symmetric stretch, ($1'$) symmetric stretch, and (2) bending.
Here,  $\tilde{\zeta_s}$ denotes the normalized S-B coupling strength, and $\gamma_s$ denotes the inverse correlation time of the bath fluctuations,  $V_{\mathrm{SL}}^{(s)}$ 
denotes the SL interaction, and $\tilde{g}_{s^3}$ is the cubic anharmonicity for the $s$ vibrational mode, respectively.  Since the LL interaction does not play a significant role in high-frequency intramolecular modes, we set $V_{\mathrm{LL}}^{(s)} = 0$.} 
\begin{tabular}{c|ccccccc}
  \hline 
  \hline
  & s & $\omega_s$ (cm$^{-1}$)  & $\gamma_s/\omega_0$ & $\tilde{\zeta}_s$ &  ${V}_{SL}^{(s)}$  & $\tilde{g}_{s^3}$ \\
\hline
\multirow{3}{*}{Flexible SPC/E } 
&1 &  $3202$ & $8.07\times 10^{1}$ &  $4.91\times 10^{-5}$  & $1.00$&$-4.91\times 10^{-3}$ \\
&$1'$ & $3123$ & $7.87\times 10^{1}$ &$3.74\times 10^{-5}$ & $1.00$ & $-5.31\times 10^{-3}$ \\
&2 &  $1592$ & $4.01\times 10^{1}$ &  $9.28\times 10^{-6}$ &  $1.00$  & $-4.12\times 10^{-2}$ \\
\hline
\multirow{3}{*}{Ferguson } 
& 1 &  $3513$ & $8.86\times 10^{1}$ &$4.96\times 10^{-5}$  &  $1.00$&$-3.33\times 10^{-3}$\\
& $1'$ & $3413$ & $8.60\times 10^{1}$ & $3.37\times 10^{-5}$&  $1.00$ &$-4.62\times 10^{-3}$\\
& 2 & $1636$ & $4.12\times 10^{1}$ & $1.07\times 10^{-4}$& $1.00$  & $-4.28\times 10^{-2}$\\
\bottomrule
  \hline \hline
\end{tabular}
\end{table*}

\begin{table*}[!htb]
\caption{\label{tab:spce_drude_3bo_mode}
Optimized mode–mode coupling parameters of the MAB model trained from flexible SPC/E and Ferguson potentials with the Drude SDF and SL interaction for (1) anti-symmetric stretch, ($1'$) symmetric stretch, and (2) bending modes.} 
\begin{tabular}{c|cccc}
  \hline \hline
& $\mathrm{s-s'}$ & $\tilde{g}_{ss'}$  & $\tilde{g}_{s^2s'}$ & $\tilde{g}_{s{s'}^2}$  \\
 \hline
 \multirow{3}{*}{Flexible SPC/E }
 &  $\mathrm{1-1'}$ & $5.67\times 10^{-1}$ & $3.23\times 10^{-3}$ & $3.44\times 10^{-3}$\\
 & $\mathrm{1-2}$ & $1.21\times 10^{0}$ & $9.22\times 10^{-3}$ & $1.96\times 10^{-2}$  \\
 &  $\mathrm{1'-2}$ &$1.13\times 10^{0}$ & $7.05\times 10^{-3}$ & $1.91\times 10^{-2}$\\
 \hline
 \multirow{3}{*}{Ferguson}
 & $\mathrm{1-1'}$ & $4.77\times 10^{-1}$ & $2.46\times 10^{-3}$ & $1.87\times 10^{-3}$ \\
 & $\mathrm{1-2}$ &  $1.06\times 10^{0}$ & $6.76\times 10^{-3}$ & $1.65\times 10^{-2}$ \\
 & $\mathrm{1'-2}$ &$1.04\times 10^{0}$ & $7.81\times 10^{-3}$ & $1.69\times 10^{-2}$\\
\hline \hline
\end{tabular}
\end{table*}

\begin{table*}[!htb]
\caption{\label{tab:spce_drude_3bo_compact_products}
Optimized inter-molecular $\alpha$th PBM paramters of the MAB model trained from flexible SPC/E and Ferguson potentials. Frequencies are \(\omega_\alpha\) in cm\(^{-1}\) with \(\omega_0=4000~\mathrm{cm}^{-1}\)}. 
\centering
\begin{tabular}{c|ccccccccc}
  \hline \hline
& $\alpha$ & $\omega_\alpha$ (cm$^{-1}$) & $\gamma_\alpha/\omega_0$
& $\tilde{\zeta}_1^\alpha \tilde{V}_{\rm LL}$ & $\tilde{\zeta}_1^\alpha \tilde{V}_{\rm SL}$
& $\tilde{\zeta}_{1'}^\alpha \tilde{V}_{\rm LL}$ & $\tilde{\zeta}_{1'}^\alpha \tilde{V}_{\rm SL}$
& $\tilde{\zeta}_2^\alpha \tilde{V}_{\rm LL}$ & $\tilde{\zeta}_2^\alpha \tilde{V}_{\rm SL}$ \\
 \hline
 \multirow{3}{*}{Flexible SPC/E }
  &4 & 500 & $1.38\times10^{-2}$
  & $3.47\times 10^{-9}$ & $8.96\times 10^{-5}$
  & $3.94\times 10^{-9}$ & $1.19\times 10^{-4}$
  & $5.85\times 10^{-9}$ & $1.05\times 10^{-4}$ \\
&5 & 720 & $1.38\times10^{-2}$
  & $2.47\times 10^{-7}$ & $7.46\times 10^{-5}$
  & $2.56\times 10^{-7}$ & $8.37\times 10^{-5}$
  & $4.85\times 10^{-7}$ & $1.13\times 10^{-4}$ \\
&6 & 804 & $1.38\times10^{-2}$
  & $2.76\times 10^{-7}$ & $7.83\times 10^{-5}$
  & $2.57\times 10^{-7}$ & $7.42\times 10^{-5}$
  & $4.40\times 10^{-7}$ & $1.35\times 10^{-4}$ \\
   \hline
 \multirow{3}{*}{Ferguson}
&4 & 476 & $1.38\times10^{-2}$
  & $2.86\times 10^{-9}$ & $6.88\times 10^{-5}$
  & $2.17\times 10^{-9}$ & $7.12\times 10^{-5}$
  & $3.59\times 10^{-7}$ & $1.77\times 10^{-4}$ \\
&5 & 604 & $1.38\times10^{-2}$
  & $2.08\times 10^{-9}$ & $6.81\times 10^{-5}$
  & $1.92\times 10^{-9}$ & $6.11\times 10^{-5}$
  & $3.75\times 10^{-7}$ & $1.83\times 10^{-4}$ \\
&6 & 692 & $1.38\times10^{-2}$
  & $2.34\times 10^{-9}$ & $7.84\times 10^{-5}$
  & $1.52\times 10^{-9}$ & $5.85\times 10^{-5}$
  & $3.36\times 10^{-7}$ & $1.63\times 10^{-4}$ \\
\bottomrule
\end{tabular}
\end{table*}

\section{CHFPE-2DVS}
\label{HEOM_Drude}.
For the MBA model [Eqs. \eqref{eqn:H_total}–\eqref{eqn:inter}] with the Drude spectral density [Eq. \eqref{eq:drude}], QHFPE have been formulated to describe two vibrational modes, encompassing both intramolecular and intermolecular dynamics.\cite{TT23JCP1} In parallel, CHFPE-2DVS have been developed to treat three vibrational modes.\cite{HT25JCP1} Computational implementations of both approaches are publicly available.\cite{TT23JCP2,HT25JCP2}  

Here, we employ the CHFPE-2DVS developed for the MAB model. In the CHFPE, the terms associated with the Matsubara frequencies vanish, and the equations are expressed as follows\cite{IT16JCP,HT25JCP1,HT25JCP2}
\begin{eqnarray}
\label{eq:clHEOM}
  \frac{ \partial{W^{(\bm{n})}(\bm{q}, \bm{p}; t)}}{\partial t}& = 
	 (\hat{L}(\bm{q}, \bm{p}) -\sum_{s} n_{s} \gamma_{s}) W^{(\bm{n})}(\bm{q}, \bm{p}; t) \nonumber \\
	& +\sum_{s}\hat{\Phi}_{s} W^{(\bm{n}+\bm{e}_{s})}(\bm{q}, \bm{p}; t)\nonumber \\
	&+\sum_{s}\hat{\Theta}_{s} W^{(\bm{n}-\bm{e}_{s})}(\bm{q}, \bm{p}; t),
\end{eqnarray}
where $W^{(\bm{n})}(\bm{q}, \bm{p}; t)$ is the Wigner distribution function (WDF) and the hierarchical elements are denoted as $\bm{n} = (n_1, n_2, n_3)$, where each $n_s$ is a non-negative integer representing the $s$th mode, and $\bm{e}_s$ is the unit vector in the $s$th direction. 
Note that $W^{(\bm{n})}(\bm{q}, \bm{p}; t)$ has physical meaning only when $\bm{n} = (0, 0, 0)$; for all other values of $\bm{n}$, it serves as an auxiliary WDF that accounts for non-perturbative and non-Markovian S-B interactions.\cite{T06JPSJ,T20JCP}

The classical Liouvillian $\hat{L}$ corresponding to the system Hamiltonian $H_{\rm sys}(\bm{q}, \bm{p}) \equiv \sum_{s} {H}_{A}^{(s)} + \sum_{s<s'} U_{ss'}\qty(\hat{q}_s, \hat{q}_{s'})$ is defined as
\begin{align}
\label{eq:CL_liouville}
\hat{L}(\bm{q}, \bm{p})  W(\bm{q}, \bm{p}) &\equiv \{ H_{\rm sys}(\bm{q}, \bm{p}) , W(\bm{q}, \bm{p})  \}_{\mathrm{PB}}, 
\end{align}
where $\{ \cdot, \cdot \}_{\mathrm{PB}}$ denotes the Poisson bracket, given by
\begin{align}
\{ A, B \}_{\mathrm{PB}} \equiv \sum_s \left( \frac{\partial A}{\partial q_s} \frac{\partial B}{\partial p_s} - \frac{\partial A}{\partial p_s} \frac{\partial B}{\partial q_s} \right)
\end{align}
for any functions $A$ and $B$.

The operators $\hat{\Phi}_s$ and $\hat{\Theta}_s$ describe the energy exchange between the $s$th mode and its corresponding bath, and are defined as follows:\cite{TS20JPSJ,KT04JCP}
\begin{align}
\label{eq:Phi}
\hat{\Phi}_s = \frac{\partial V_s(q_s)}{\partial q_s} \frac{\partial}{\partial p_s}, 
\end{align}
and
\begin{align}
\label{eq:Theta}
\hat{\Theta}_s = \frac{m_s \zeta_s \gamma_s}{\beta} \frac{\partial V_s(q_s)}{\partial q_s} \frac{\partial}{\partial p_s}
+ \zeta_s \gamma_s p_s \frac{\partial V_s(q_s)}{\partial q_s},
\end{align}
where $\zeta_s$ is the system–bath coupling strength, $\gamma_s$ is the inverse correlation time, and $T$ is the temperature.

\section{Optimized Parameters of the MAB-Drude Model for the flexible SPC/E and Ferguson potentials}
\label{SPCE_PARA}
The optimized MAB–Drude parameters obtained from MD trajectories generated using the SPC/E and Ferguson potentials are listed in Tables \ref{tab:spce_drude_3bo_bath}, \ref{tab:spce_drude_3bo_mode}, and \ref{tab:spce_drude_3bo_compact_products}.

\bibliography{tanimura_publist.bib,TT23.bib,HT24.bib,PJT25.bib}

\end{document}